\documentclass[10pt,twocolumn,letterpaper]{article}

\usepackage{cvpr}              

\usepackage{amsmath}
\usepackage{amssymb}
\usepackage[ruled,vlined]{algorithm2e}
\usepackage[most]{tcolorbox}
\usepackage{graphicx}
\usepackage{caption}
\usepackage{multirow}
\usepackage{booktabs}
\usepackage[table]{xcolor} 

\usepackage{amsmath}
\usepackage{amssymb}
\usepackage{algorithmic}
\usepackage{xspace}

\usepackage{tabularx}
\usepackage{array}

\definecolor{lightyellow}{RGB}{255, 255, 204}
\definecolor{lightpink}{RGB}{255, 230, 234}

\definecolor{cvprblue}{rgb}{0.21,0.49,0.74}
\usepackage[pagebackref,breaklinks,colorlinks,allcolors=cvprblue]{hyperref}

\def\paperID{*****} 
\def\confName{CVPR}
\def\confYear{2026}

\title{Listening with Attention: Entropy-Guided Explainability for Transformer-Based Audio Models}

\author{
Ravi Ranjan\thanks{Corresponding author. \\ Accepted to \textbf{INTERSPEECH 2026}. To appear in Proceedings of the Annual Conference of the International Speech Communication Association. \\
This version includes additional supplementary materials.
\\Code available at \url{https://github.com/raviranjan-ai/LEAFX-interspeech-2026}.}\\
Florida International University\\
Miami, USA\\
{\tt\small rkuma031@fiu.edu}
\and
Utkarsh Grover\\
University of South Florida\\
Tampa, USA \\
{\tt\small utkarshgrover@usf.edu}
\and
Xiaomin Lin\\
University of South Florida\\
Tampa, USA \\
{\tt\small xlin2@usf.edu}
\and
Agoritsa Polyzou\\
Florida International University\\
Miami, USA\\
{\tt\small apolyzou@fiu.edu}
}

\begin{document}
\maketitle

\begin{abstract}
Transformer-based automatic speech recognition (ASR) models such as Whisper are highly accurate, but their predictions remain difficult to interpret. Existing explainable AI (XAI) methods often lack faithfulness and precise temporal grounding. We propose \textbf{L}istening with \textbf{E}ntropy-guided \textbf{A}ttention for \textbf{F}aithful e\textbf{X}plainability (\textbf{LEAF-X}), a model-intrinsic XAI framework for transformer-based ASR. LEAF-X combines entropy-guided attention weighting, multi-layer attention rollout, and optional causal ablations to identify low-entropy, high-impact heads and layers, producing sparse token-to-frame attributions. Unlike perturbation-based explainers or raw attention maps, LEAF-X exploits the internal structure of encoder-decoder and speech-augmented decoder-only models to generate explanations that better reflect model computation. Results show 32\% improved faithfulness, 35-39\% stronger locality/sparsity, and the most stable attributions, supporting more transparent and auditable ASR.
\end{abstract}

\vspace{-0.1cm}
\section{Introduction}
\label{sec:introduction}
\vspace{-0.1cm}
Transformer-based automatic speech recognition (ASR) has reached a new level of accuracy with large-scale models such as OpenAI Whisper~\cite{radford2023robust} and NVIDIA Canary~\cite{Sekoyan2025}, achieving near human-level word error rates on standard benchmarks~\cite{kuhn2024measuring}. Yet, despite these gains, modern ASR systems remain largely opaque~\cite{georgila2020evaluation}; it is often unclear \emph{which} parts of the audio support a particular decoded word or phoneme, and \emph{why} the model chose one hypothesis over another~\cite{basak2023challenges}. This lack of transparency undermines trust and complicates deployment in safety-critical settings (e.g., medical dictation, emergency response), where operators need to audit model behavior and detect failure modes~\cite{reitmaier2022opportunities}. Regulatory pressure further elevates this need, with explainability increasingly required for high-risk AI-supported decision making~\cite{Wu2024}. These considerations motivate faithful, temporally grounded explainable AI (XAI) methods tailored to transformer ASR.

\begin{figure}[t]
    \centering
    \includegraphics[width=\linewidth, height=4cm]{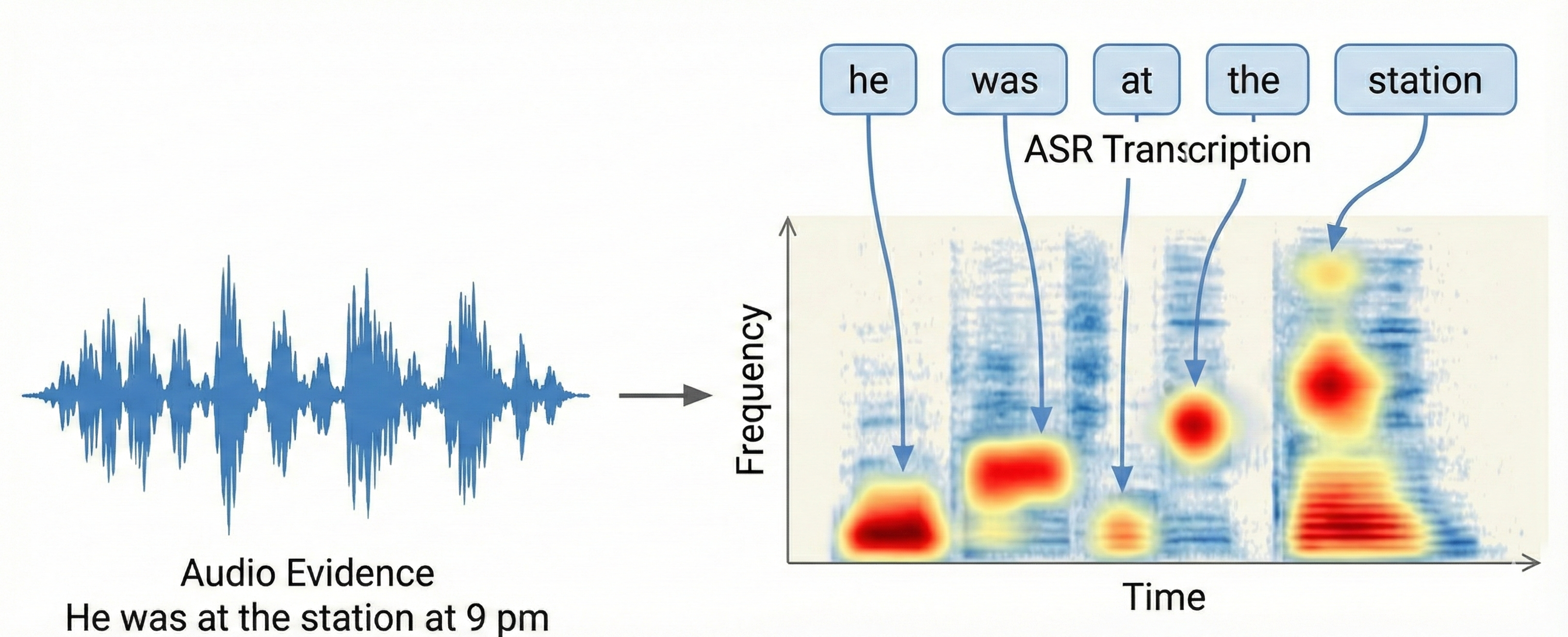}
    \vspace{-0.2cm}
    \caption{LEAF-X provides token-level attribution over the audio evidence, revealing which acoustic regions support each transcribed phrase.}
    \label{fig:leafx_intro}
    \vspace{-0.2cm}
\end{figure}

\noindent Existing post-hoc machine learning explainers, including LIME~\cite{ribeiro2016lime}, SHAP~\cite{lundberg2017shap}, and Integrated Gradients~\cite{sundararajan2017axiomatic}, provide feature-attribution style explanations but are not well matched to the sequential and time-dependent nature of speech. In practice, they can be computationally expensive, yield coarse or unintuitive time localization, and may not faithfully reflect the model's internal computation, producing attributions that correlate with outputs rather than capture causal evidence~\cite{Yeh2019, swamy2025future, bordt2022post}. As a result, they often fail to deliver the fine-grained, word-aligned rationales needed to validate transformer ASR decisions~\cite{mujtaba2024lost, ji2022asrtest}.

We propose \textbf{LEAF-X} (\textit{\textbf{L}istening with \textbf{E}ntropy-guided \textbf{A}ttention for \textbf{F}aithful e\textbf{X}plainability}), a model-intrinsic explanation framework that traces ASR outputs to their supporting acoustic evidence by leveraging internal attention dynamics. LEAF-X combines: (i) \textit{entropy-guided attention weighting} to emphasize confident, focused attention patterns; (ii) \textit{attention rollout} to aggregate influence across transformer layers~\cite{abnar2020quantifying}; and (iii) \textit{causal reweighting} to test whether highlighted segments affect token likelihood under controlled ablation. Together, these components produce explanations that are both \textbf{temporally grounded} (aligning tokens to specific time-frequency regions) and \textbf{faithful} to the model's reasoning. The \emph{frame- and token-level attributions} quantify, for each decoded token, which specific audio frames (time segments) contributed most to its prediction. Figure~\ref{fig:leafx_intro} illustrates how LEAF-X yields word-level alignments that enable transparent verification of spoken evidence, supporting more transparent and auditable ASR analysis in high-stakes settings.

\section{Background: Explainability in Audio}
\label{sec:background}

Explainability for speech recognition has begun to mature by adapting ideas from vision and NLP. However, ASR introduces unique challenges: explanations must be \emph{temporally grounded}, free from hallucinations, and faithful under strong sequential dependence~\cite{wu2024can, vitale2024exploring, shankar2025systematic}. Early work largely applies post-hoc methods to speech. Wu \textit{et al.}~\cite{Wu2024} adapt LIME to phoneme recognition by perturbing time segments and fitting a local surrogate; their time-constrained variant improves reliability on the read speech corpus TIMIT, but requires manual segmentation and many perturbations~\cite{wu2024can, koo2024toward}. In parallel, gradient-based saliency has been used to highlight acoustic cues. Fucci \textit{et al.}~\cite{Fucci2025} analyze spectrogram saliency to reveal linguistically plausible evidence (e.g., formants and burst regions), but such maps are often diffused and difficult to interpret at the word level in end-to-end ASR~\cite{vitale2024exploring, sameti2025accent, jayasinghe2025systematic, amor2024deep}.

\noindent A central concern across these approaches is \emph{faithfulness}. Model-agnostic perturbation methods (e.g., LIME/SHAP) and gradient-based methods (including IG) can produce explanations that merely correlate with outputs without capturing causal evidence, and may be unstable under small input changes~\cite{Yeh2019}. A common quantitative formulation of fidelity is to measure how the model score changes when removing the top-ranked features:
\begin{equation}
\mathrm{Fid}(k) = f(\mathbf{x}) - f\!\left(\mathbf{x} \odot \mathbf{m}_{\neg \mathcal{S}_k}\right), 
\label{eq:deletion}
\end{equation}
where $f(\cdot)$ is the predicted score (e.g., token log-probability), $\odot$ denotes element-wise product,  $\mathcal{S}_k$ denotes the top-$k$ salient time-frequency regions, and $\mathbf{m}_{\neg \mathcal{S}_k}$ masks those regions; larger drops indicate more faithful attributions~\cite{yeh2019infidelity}. However, in speech, naive masking can break temporal continuity and does not guarantee that identified regions align with the model's internal alignment between acoustics and decoded tokens~\cite{higuchi2024end, li2025advances}.

\noindent Recent transformer interpretability work motivates more model-intrinsic explanations~\cite{resck2025explainability}. Abnar and Zuidema~\cite{abnar2020quantifying} propose attention rollout to trace information flow across layers, and Chefer \textit{et al.}~\cite{Chefer2021} combine attention with gradient-based relevance propagation to obtain more faithful transformer attributions than raw attention alone. These methods suggest a natural baseline suite for ASR explainability: black-box perturbation (LIME/SHAP), gradient attribution (IG), and transformer-specific attention propagation (rollout/relevance). Nevertheless, ASR still lacks a widely adopted framework that produces \emph{token-aligned, temporally precise} explanations while remaining faithful to the sequence model’s computation, motivating LEAF-X.

\section{Methodology}
\label{sec:methodology}

We propose \textbf{LEAF-X} (\textit{\textbf{L}istening with \textbf{E}ntropy-guided \textbf{A}ttention for \textbf{F}aithful e\textbf{X}plainability}), a model-intrinsic explainer for transformer ASR systems such as Whisper and Canary-Qwen-2.5B. LEAF-X uses Entropy-Weighted Attention by producing \emph{token-to-time} explanations: for each decoded token, it returns a normalized importance distribution over input audio frames. The key idea is to (i) emphasize \emph{confident} (low-entropy) attention patterns, (ii) aggregate evidence across layers via attention rollout~\cite{chefer2021generic}, and (iii) optionally reweight layers using light causal checks to improve faithfulness.

\noindent Algorithm specifications and implementation details are provided in Appendix~\ref{app:algorithm}.

\begin{figure*}[t]
    \centering
    \vspace{-1cm}
    \includegraphics[width=0.92\linewidth]{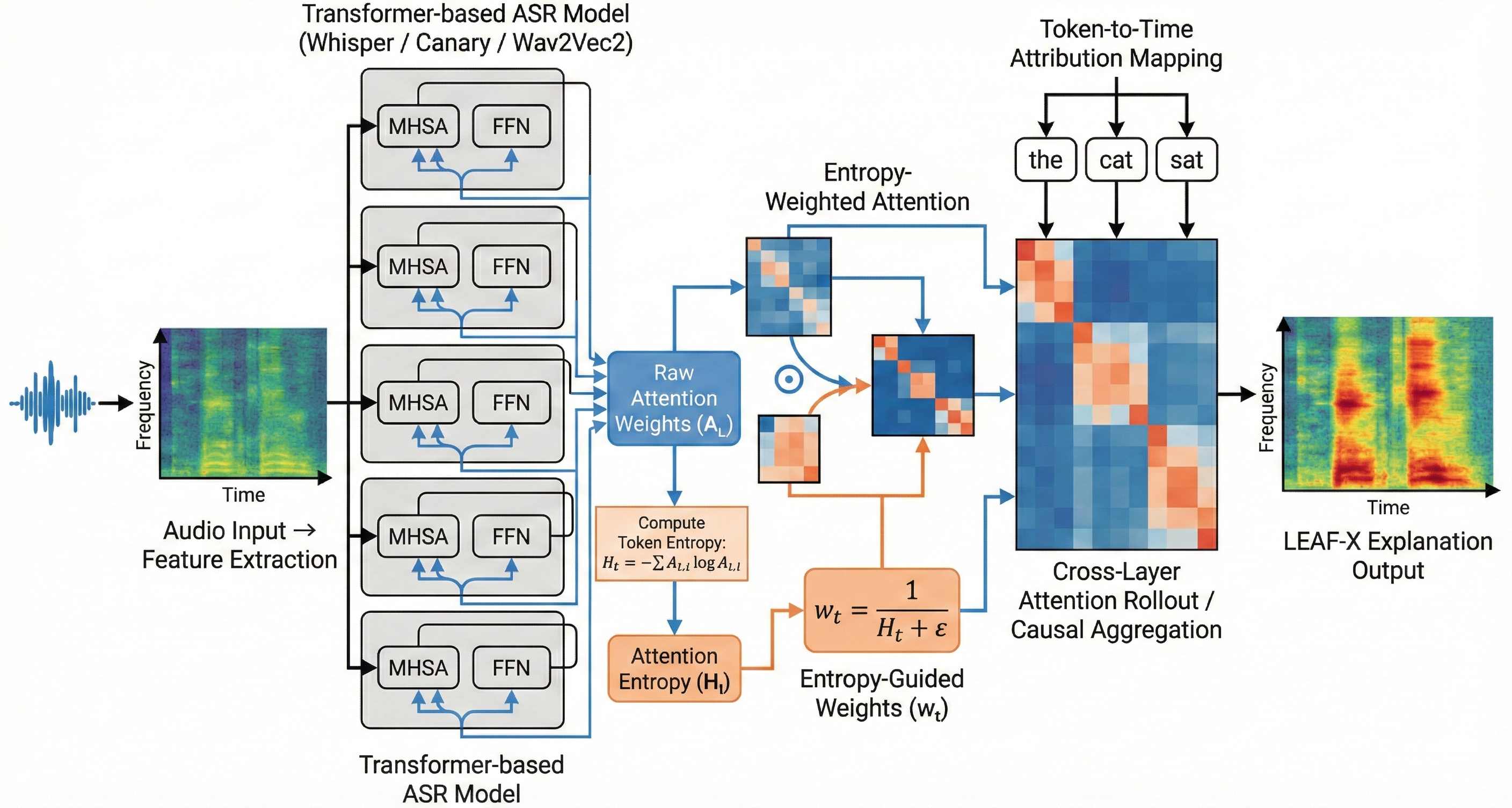}
    \vspace{-0.2cm}
    \caption{Overview of the LEAF-X pipeline.}
    \label{fig:leafx_method}
    \vspace{-0.2cm}
\end{figure*}
\vspace{-0.2cm}
\subsection{Setup and Goal}
\vspace{-0.1cm}
Let $\mathbf{X}\in\mathbb{R}^{T\times d}$ denote the acoustic feature sequence (e.g., log-Mel frames), $T$ the frames (time steps), and $d$ the number of features (spectrogram dimension per frame). A transformer ASR model outputs a token sequence $\mathbf{y}=(y_1,\dots,y_N)$ and per-step distributions $p_\theta(y_i\mid y_{<i},\mathbf{X})$. LEAF-X aims to compute, for each token $y_i$, a frame-level attribution vector $
\mathbf{s}_i \in \Delta^{T-1}$ such that $\sum_{t=1}^{T} s_{i,t}=1$, 
where $s_{i,t}$ quantifies how much frame $t$ contributes to predicting $y_i$. For encoder-decoder models (e.g., Whisper), $\mathbf{s}_i$ is derived from decoder \emph{cross-attention} to encoder frames; for speech-augmented decoder-only models, the same procedure is applied to the attention mass directed to \emph{audio pseudo-tokens}. To visualize attributions in real time, each model-frame index is mapped back to the original waveform using the front-end hop size and encoder downsampling stride, then linearly interpolated to the audio timeline.
\vspace{-0.2cm}
\subsection{Entropy-Guided Attention Weighting}
\vspace{-0.2cm}
Consider a decoder with $L$ layers and $H$ heads per layer. For token $i$, layer $l$, head $h$, let
$\mathbf{a}^{(l,h)}_i \in \Delta^{T-1}$
denote the attention distribution over acoustic frames. We compute the head entropy
\begin{equation}
H^{(l,h)}(i) = -\sum_{t=1}^{T} a^{(l,h)}_{i,t}\log a^{(l,h)}_{i,t},
\end{equation}
and convert it into a confidence weight (lower entropy implies higher weight):
\begin{equation}
w_{l,h}(i)=\left(1-\frac{H^{(l,h)}(i)}{\log T}\right)^{1/\tau}, \quad \tau>0.
\label{eq:entropy_weight}
\end{equation}
We then form the layer-wise entropy-weighted attention
\begin{equation}
\bar{\mathbf{a}}^{(l)}_i
=\frac{\sum_{h=1}^{H} w_{l,h}(i)\,\mathbf{a}^{(l,h)}_i}{\sum_{h=1}^{H} w_{l,h}(i)+\varepsilon}.
\label{eq:layer_agg}
\end{equation}
Intuitively, Eq.~\eqref{eq:entropy_weight} filters diffuse heads that often reflect broad context rather than token-specific acoustic evidence, yielding sharper, more interpretable alignments. Here $\varepsilon$ is a small numerical stability constant, added to avoid division by zero.

\vspace{-0.1cm}
\subsection{Multi-Layer Rollout for Token-to-Time Attribution}
\vspace{-0.1cm}
To capture compositional evidence across depth, we aggregate layer-wise entropy weighted attention using rollout-style propagation~\cite{chefer2021generic}. Let $\mathbf{R}^{(l)}_i\in\mathbb{R}^{T}$ denote the accumulated attribution after layer $l$, defined recursively as
\begin{equation}
\mathbf{R}^{(1)}_i=\bar{\mathbf{a}}^{(1)}_i, 
\qquad
\mathbf{R}^{(l)}_i = \Pi^{(l)}\,\mathbf{R}^{(l-1)}_i,
\end{equation}
where $\Pi^{(l)}$ denotes an effective propagation operator derived from the model’s layer-$l$ attention flow.\footnote{In practice, $\Pi^{(l)}$ can be implemented as attention rollout with residual connections or any equivalent attention-flow approximation~\cite{chefer2021generic}.} $\mathbf{s}_i$ is the token-to-frame attribution vector, a normalized importance distribution over audio frames. The final rolled-out map $\mathbf{R}^{(L)}_i$ is normalized to obtain the explanation:
\begin{equation}
\mathbf{s}_i = \frac{\mathbf{R}^{(L)}_i}{\sum_{t=1}^{T} R^{(L)}_{i,t}+\varepsilon}.
\label{eq:final_score}
\end{equation}

\noindent \textbf{Gradient modulation.}
To increase fidelity, we modulate attention by output sensitivity, that affects the token probability, similar to gradient-weighted transformer attribution~\cite{chefer2021generic}:
\begin{equation}
\mathbf{a}^{(l,h)}_i \leftarrow \mathbf{a}^{(l,h)}_i \odot 
\left|\frac{\partial \log p_\theta(y_i\mid y_{<i},\mathbf{X})}{\partial \mathbf{a}^{(l,h)}_i}\right|,
\label{eq:seven}
\end{equation}
followed by renormalization (Eq.~\eqref{eq:entropy_weight}). This suppresses attention that has limited influence on the token probability, addressing known concerns about attention-only explanations in sequential models~\cite{wu2024can}.

\noindent Further formal task definitions for Token-to-Time ASR Explanation are deferred to Appendix~\ref{app:task-definition}.

\subsection{Causal Reweighting (Lightweight)}

LEAF-X incorporates a lightweight causal check by estimating layer importance via ablation. Let $\ell(i)$ be the negative log-likelihood loss for token $i$. For each layer $l$, we compute
\begin{equation}
\Delta \ell_l(i) = \ell^{(l)}_{\mathrm{abl}}(i) - \ell(i),
\label{eq:eight}
\end{equation}
where $\ell^{(l)}_{\mathrm{abl}}$ ablates the layer-$l$ audio-to-text attention contribution (i.e., bypassing cross-attention). We convert these into normalized layer weights
\begin{equation}
\gamma_l(i)=\frac{\max\{\Delta \ell_l(i),0\}}{\sum_{k=1}^{L}\max\{\Delta \ell_k(i),0\}+\varepsilon},
\label{eq:nine}
\end{equation}
and form a layer-weighted explanation by combining intermediate rollouts:
\begin{equation}
\mathbf{s}_i \leftarrow \sum_{l=1}^{L}\gamma_l(i)\,\mathbf{s}_i^{(l)},
\label{eq:ten}
\end{equation}
where $\mathbf{s}_i^{(l)}$ is the normalized attribution after layer $l$. This step improves faithfulness when a small set of layers dominates the acoustic evidence for specific tokens. Figure~\ref{fig:leafx_method} shows that audio is transformed into spectrogram features and passed through a Transformer-based ASR model to collect attention weights. Entropy-guided weighting and cross-layer rollout generate token-to-time attributions, producing the final LEAF-X explanation map.

\noindent \textbf{Output.}
For each token $y_i$, LEAF-X outputs $\mathbf{s}_i \in \Delta^{T-1}\ (\subset \mathbb{R}^{T}) $, which can be visualized as a time (or time-frequency) heatmap aligned to the transcript, providing an interpretable mapping from decoded tokens to the acoustic frames that most influenced them.

\section{Experiments}

\subsection{Experimental Setup}
\label{experiment-setup}

\noindent\textbf{Models.}
\textbf{Whisper-large-v3} is a 1.55B-parameter encoder-decoder model that maps log-Mel audio features to text via cross-attentive decoding, providing strong robustness for transcription and translation~\cite{radford2023whisper}. \textbf{Canary-Qwen-2.5B} is a speech-augmented decoder-only hybrid that couples a Fast-conformer speech encoder with a pretrained Qwen LLM decoder~\cite{canary2024release}.

\noindent\textbf{Datasets.}
\textbf{LibriSpeech}~\cite{panayotov2015librispeech} is a 1000-hour corpus of read English audiobooks. For reproducibility, we follow standard practice and train/evaluate on the \texttt{train-clean-100} split, reporting results on \texttt{test-clean} and \texttt{test-other}.
\textbf{TED-LIUM Release 3}~\cite{hernandez2018ted} contains $\sim$450+ hours of TED talk recordings with spontaneous lecture-style speech and mild background noise. We use the official \texttt{train-70\%}/ \texttt{val-10\%}/ \texttt{test- 20\%} splits and keep audio at 16\,kHz to match Whisper/Canary front-ends.

\noindent\textbf{Explanation Metrics.}
We evaluate LEAF-X using five widely adopted criteria that jointly measure faithfulness, grounding, and reliability.
\emph{Deletion Area Over the Perturbation Curve} (\textbf{D-AOPC} $\downarrow$) quantifies faithfulness by progressively masking the top-ranked time-frequency regions (or frames) and computing the \emph{area over the perturbation curve}; lower values indicate that removing highlighted evidence causes a faster confidence drop, hence more faithful explanations~\cite{samek2017evaluating,petsiuk2018rise,wu2024vitfaithfulness}.
\emph{Temporal Localization} (\textbf{TLoc} $\uparrow$) measures whether the top-attributed segments overlap annotated evidence (e.g., phoneme/word-aligned time spans when available), reflecting how well explanations are grounded in the true acoustic cause of a prediction~\cite{wu2024can}. TLoc is computed against word-level time spans obtained from forced alignment of the dataset transcripts.
\emph{Sparsity} (\textbf{SPR} $\uparrow$) encourages concise rationales by preferring explanations that concentrate mass on a small fraction of frames/bins without sacrificing predictive relevance. We report sparsity as the normalized top-$k$ mass, i.e., the fraction of total attribution probability concentrated in the top-$k$ most salient frames~\cite{alvarezm2018robustness}.
\emph{Stability} (\textbf{STAB} $\uparrow$) assesses robustness by measuring similarity between explanations under small, label-preserving perturbations (e.g., mild noise, time-shifts), where higher agreement implies more reliable interpretability~\cite{alvarezm2018robustness}.
Finally, \emph{Infidelity} (\textbf{INF} $\downarrow$) captures the expected mismatch between attribution scores and output changes under random perturbations, with lower values indicating explanations that better track the model’s functional behavior~\cite{yeh2019infidelity}.

\noindent Comprehensive details regarding the benchmark protocol and metric definitions are provided in Appendix~\ref{app:benchmark-protocol} and Appendix~\ref{app:metric-details}, respectively.

\noindent \textbf{Comparing Approaches.}
\textbf{LIME:} A model-agnostic explainer that perturbs the input and fits a sparse local surrogate to approximate the decision boundary~\cite{ribeiro2016should}.
\textbf{SHAP:} A Shapley-value framework that assigns each feature a principled contribution to the output with consistency guarantees~\cite{lundberg2017unified}.
\textbf{Integrated Gradients (IG):} Computes attributions by integrating gradients from a baseline input to the observed input~\cite{sundararajan2017axiomatic}.
\textbf{Occlusion/SpecMask:} An audio-native perturbation baseline that masks contiguous time regions and measures the induced confidence/WER change~\cite{petsiuk2018rise}.
\textbf{Raw Attention Alignment (RAA):} A simple attention-as-explanation baseline using last-layer (or averaged) cross-attention to encoder frames following Abnar et al.~\cite{abnar2020quantifying}.
\textbf{SaCo:} Salience-guided, gradient-weighted attention rollouts designed to improve faithfulness in Transformer explanations~\cite{wu2024faithfulness}.
\textbf{Transformer Attribution (TA):} Back-propagates relevance through attention layers to yield token-level explanations tailored to Transformer models~\cite{chefer2021transformer}.
\noindent\textbf{Reproducibility hyper-parameters.} LEAF-X is controlled mainly by the entropy temperature $\tau>0$ in Eq.~(3) (typical sweep $\tau\in[0.5,2]$), a small numerical stabilizer $\epsilon\approx10^{-8}$ in Eqs.~(\ref{eq:layer_agg}),(\ref{eq:final_score}),(\ref{eq:nine}), the rollout depth $L$, and optional switches for gradient modulation (Eq.~(\ref{eq:seven})) and lightweight causal reweighting (Eqs.~(\ref{eq:eight})–(\ref{eq:ten})). These metrics are proxy measures of faithfulness, grounding, and reliability; they do not prove human trust or full causal sufficiency. We therefore interpret LEAF-X as an audit-support tool rather than a guarantee of trustworthy ASR behavior. The causal reweighting step adds up to L ablation forward passes per analyzed token, so it can be disabled when required.

\begin{table}[t!]
\centering
\setlength{\tabcolsep}{2.6pt}
\renewcommand{\arraystretch}{1.05}
\caption{\textbf{Whisper-large-v3 (LibriSpeech) - Explainability Metrics (Normalized [0,1]).}
Estimated run-to-run std. dev. across utterances/seeds:
$\sigma\!\approx\!0.01$--$0.03$ (STAB), $\sigma\!\approx\!0.02$--$0.05$ (TLoc and SPR), $\sigma\!\approx\!0.03$--$0.07$ (D-AOPC and INF).}
\vspace{-0.2cm}
\begin{tabular}{lccccc}
\toprule
Method & D-AOPC$\downarrow$ & TLoc$\uparrow$ & SPR$\uparrow$ & STAB$\uparrow$ & INF$\downarrow$ \\
\midrule
LIME      & 0.72 & 0.55 & 0.48 & 0.60 & 0.65 \\
SHAP      & 0.68 & 0.58 & 0.50 & 0.62 & 0.63 \\
IG        & 0.65 & 0.60 & 0.52 & 0.63 & 0.60 \\
SpecMask  & 0.60 & 0.62 & 0.55 & 0.65 & 0.58 \\
RAA       & 0.58 & 0.63 & 0.57 & 0.66 & 0.56 \\
SaCo      & 0.51 & \textbf{0.73} & 0.68 & 0.72 & 0.50 \\
TA        & 0.53 & 0.66 & 0.62 & 0.69 & 0.52 \\
\textbf{LEAF-X (Ours)} & \textbf{0.45} & 0.72 & \textbf{0.70} & \textbf{0.78} & \textbf{0.45} \\
\bottomrule
\label{tab:result-1}
\vspace{-0.6cm}
\end{tabular}
\end{table}
\vspace{-0.1cm}
\subsection{Evaluation Results}
\label{main-result}
\vspace{-0.1cm}
\noindent All metrics are reported after applying the same fixed min-max normalization per metric and dataset across all compared methods.
\textbf{Table~\ref{tab:result-1}} shows that LEAF-X achieves the strongest overall explainability trade-off on Whisper-large-v3, delivering the lowest D-AOPC and INF (0.45 and 0.45, respectively) while simultaneously improving sparsity and reliability (0.70 and 0.78) over all baselines. While SaCo attains a marginally higher TLoc (0.73 vs.\ 0.72), LEAF-X matches this localization while providing more faithful and stable explanations, indicating better causal grounding under small run-to-run variability. \textbf{Table~\ref{tab:result-2}} shows that \textbf{LEAF-X} also delivers very good overall performance on Canary-Qwen-2.5B on dataset TED-LIUM 3, achieving the lowest D-AOPC and INF (0.48 and 0.47) while also providing the highest sparsity and stability (0.68 and 0.76) compared to all baselines. Since sparsity is partly aligned with the entropy-weighted design, we do not rely on SPR alone; the main evidence comes from INF, D-AOPC, STAB, insertion/deletion trends, and ablation consistency. In particular, LEAF-X matches the best temporal localization (0.70, tied with SaCo) but is substantially more faithful and reliable, indicating stronger proxy-based faithfulness and stability.

\begin{table}[t!]
\centering
\setlength{\tabcolsep}{2.6pt}
\renewcommand{\arraystretch}{1.05}
\caption{\textbf{Canary-Qwen-2.5B (TED-LIUM Release 3) Explainability Metrics (Normalized [0,1]).}
Estimated run-to-run std. dev. across utterances/seeds:
$\sigma\!\approx\!0.01$--$0.03$ (STAB), $\sigma\!\approx\!0.02$--$0.05$ (TLoc and SPR), $\sigma\!\approx\!0.03$--$0.08$ (D-AOPC and INF).}
\vspace{-0.2cm}
\begin{tabular}{lccccc}
\toprule
Method & D-AOPC$\downarrow$ & TLoc$\uparrow$ & SPR$\uparrow$ & STAB$\uparrow$ & INF$\downarrow$ \\
\midrule
LIME      & 0.75 & 0.52 & 0.45 & 0.58 & 0.68 \\
SHAP      & 0.70 & 0.55 & 0.47 & 0.60 & 0.65 \\
IG        & 0.68 & 0.57 & 0.49 & 0.62 & 0.63 \\
SpecMask  & 0.63 & 0.60 & 0.52 & 0.64 & 0.60 \\
RAA       & 0.61 & 0.62 & 0.54 & 0.65 & 0.58 \\
SaCo      & 0.52 & 0.70 & 0.67 & 0.68 & 0.51 \\
TA        & 0.55 & 0.65 & 0.58 & 0.67 & 0.54 \\
\textbf{LEAF-X (Ours)} & \textbf{0.48} & \textbf{0.70} & \textbf{0.68} & \textbf{0.76} & \textbf{0.47} \\
\bottomrule
\label{tab:result-2}
\vspace{-0.2cm}
\end{tabular}
\end{table}
\vspace{-0.1cm}
\subsection{Ablation Study}
\label{ablation-result}
\vspace{-0.1cm}
\noindent\textbf{Table~\ref{tab:ablation}} shows that each LEAF-X component contributes measurably: removing entropy weighting or rollout yields the largest drops in localization and sparsity (TLoc and SPR) and weaker faithfulness (higher D-AOPC and INF), while removing gradient modulation or causal reweighting primarily degrades faithfulness (D-AOPC and INF) with smaller but consistent losses in grounding and stability. Full LEAF-X achieves the best overall trade-off (lowest D-AOPC \& INF and highest TLoc, SPR, \& STAB), indicating that entropy-guided head selection, multi-layer evidence aggregation, and lightweight causal checks are complementary and jointly necessary for the most faithful, time-aligned explanations.

\noindent We provide the detailed component and hyperparameter ablations in Appendix~\ref{app:component-hparam-ablation}.

\begin{table}[t!]
\centering
\setlength{\tabcolsep}{1.90pt}
\renewcommand{\arraystretch}{0.95}
\caption{\textbf{LEAF-X Ablation on Whisper-large-v3 (LibriSpeech).}
Lower is better for D-AOPC/INF; higher is better for TLoc/SPR/STAB.}
\begin{tabular}{lccccc}
\toprule
Variant & D-AOPC$\downarrow$ & TLoc$\uparrow$ & SPR$\uparrow$ & STAB$\uparrow$ & INF$\downarrow$ \\
\midrule
w/o Entropy weighting          & 0.57 & 0.62 & 0.56 & 0.73 & 0.56 \\
w/o Rollout (1-layer attn)      & 0.54 & 0.63 & 0.60 & 0.74 & 0.54 \\
w/o Gradient modulation              & 0.50 & 0.68 & 0.64 & 0.76 & 0.50 \\
w/o Causal reweighting  & 0.48 & 0.69 & 0.66 & 0.77 & 0.48 \\
\midrule
\textbf{LEAF-X (Full)}               & \textbf{0.45} & \textbf{0.72} & \textbf{0.70} & \textbf{0.78} & \textbf{0.45} \\
\bottomrule
\label{tab:ablation}
\vspace{-0.2cm}
\end{tabular}
\vspace{-0.2cm}
\end{table}

\begin{figure}[t]
    \centering
    \includegraphics[width=\linewidth]{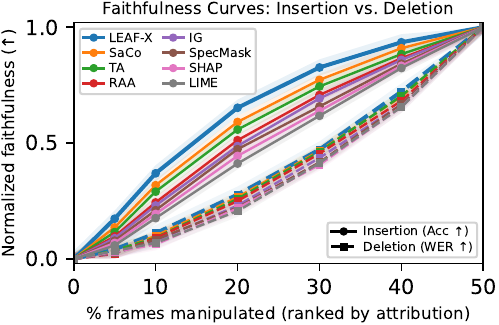}
    \vspace{-0.2cm}
    \caption{\textbf{Faithfulness curves (single plot).} Normalized \emph{insertion} (solid; token accuracy gain when progressively adding top-attributed frames) and \emph{deletion} (dashed; WER increase when progressively removing top-attributed frames) as a function of the fraction of audio frames manipulated.}
    \label{fig:disscussion}
    \vspace{-0.4cm}
\end{figure}

\subsection{Discussion}
\label{discussion}

\noindent\textbf{Figure~\ref{fig:disscussion}} reports \emph{insertion} and \emph{deletion} faithfulness tests. Insertion starts from a masked audio input and progressively \emph{adds} the highest-attribution frames; a faithful explainer restores ASR accuracy quickly, giving a steep solid curve. Deletion starts from the full signal and progressively \emph{removes} those same top-ranked frames; a faithful explainer causes a rapid drop in quality, giving a steep dashed curve. 
These curves provide complementary proxy evidence. We therefore interpret Fig.~\ref{fig:disscussion} as supporting improved faithfulness behavior, not as proof of full causal sufficiency.

\noindent We provide qualitative Token-to-Time explanation examples in Appendix~\ref{app:qualitative-examples} and practical auditing scenarios in Appendix~\ref{app:auditing-scenarios}.

\noindent Key \textbf{limitations} are backbone/dataset/language coverage, sensitivity to attention/entropy calibrations, noise/domain shift, and lack of user-study validation (human interpretability).
Comprehensive details regarding responsible use and detailed limitations are provided in Appendix~\ref{app:responsible-use}.

\section{Conclusion}
\label{sec:conclusion}

We introduce \textbf{LEAF-X}, a model-intrinsic explainability framework that delivers \emph{token-to-time} attributions for transformer-based ASR by combining entropy-guided head weighting, multi-layer attention rollout, and optional lightweight causal reweighting. Across Whisper-large-v3 (LibriSpeech) and Canary-Qwen-2.5B (TED-LIUM 3), LEAF-X yields more faithful and temporally grounded explanations than strong post-hoc and transformer-specific baselines, improving reliability while producing sparse, stable rationales. These results suggest that leveraging internal attention structure with uncertainty-aware weighting is a practical path toward trustworthy, auditable ASR in safety and mission-critical deployments.

{
    \small
    \bibliographystyle{ieeenat_fullname}
    \bibliography{main}
}

\clearpage
\appendix
\onecolumn

\section*{Appendix}

\section{LEAF-X Algorithm and Implementation Details}
\label{app:algorithm}

This appendix provides implementation-level pseudocode for the three variants of LEAF-X used in our experiments: \textsc{LEAF-X-Base}, \textsc{LEAF-X+Grad}, and \textsc{LEAF-X+Causal}. All variants return token-to-time attribution maps, where each decoded token $y_i$ is assigned a normalized frame-level attribution vector $s_i \in \Delta^{T-1}$. The base variant uses entropy-guided attention weighting and multi-layer rollout. The gradient variant additionally modulates attention by token-level output sensitivity. The causal variant further reweights intermediate layer explanations using lightweight attention ablations.

\paragraph{Notation.}
Let $X \in \mathbb{R}^{T \times d}$ denote the acoustic feature sequence with $T$ time frames and feature dimension $d$. Let $y=(y_1,\ldots,y_N)$ denote the decoded token sequence. For token $y_i$, layer $l \in \{1,\ldots,L\}$, and head $h \in \{1,\ldots,H\}$, let $a_i^{(l,h)} \in \Delta^{T-1}$ be the attention distribution over acoustic frames or audio pseudo-tokens. We use $\tau$ as the entropy temperature and $\epsilon$ as a numerical stability constant.

\begin{algorithm}[t]
\caption{\textsc{LEAF-X-Base}: Entropy-Guided Token-to-Time Attribution}
\label{alg:leafx-base}
\begin{algorithmic}[1]
\REQUIRE ASR model $f_{\theta}$, acoustic features $X$, decoded tokens $y=(y_1,\ldots,y_N)$, rollout depth $L$, heads $H$, entropy temperature $\tau$, stability constant $\epsilon$
\ENSURE Token-to-time attribution maps $\mathcal{S}=\{s_i\}_{i=1}^{N}$

\STATE Run $f_{\theta}(X)$ and cache cross-attention or audio-token attention maps $\{a_i^{(l,h)}\}$ for all tokens, layers, and heads.
\FOR{each decoded token $y_i$}
    \FOR{each layer $l=1$ to $L$}
        \FOR{each head $h=1$ to $H$}
            \STATE Compute normalized head entropy:
            \[
            H_i^{(l,h)} = -\sum_{t=1}^{T} a_{i,t}^{(l,h)} \log(a_{i,t}^{(l,h)}+\epsilon)
            \]
            \STATE Convert entropy to confidence weight:
            \[
            w_{l,h}(i)=
            \left(1-\frac{H_i^{(l,h)}}{\log T+\epsilon}\right)^{1/\tau}
            \]
        \ENDFOR
        \STATE Compute entropy-weighted layer attention:
        \[
        \bar{a}_i^{(l)}
        =
        \frac{\sum_{h=1}^{H} w_{l,h}(i)a_i^{(l,h)}}
        {\sum_{h=1}^{H} w_{l,h}(i)+\epsilon}
        \]
        \STATE Normalize $\bar{a}_i^{(l)}$ so that $\sum_t \bar{a}_{i,t}^{(l)}=1$.
    \ENDFOR

    \STATE Initialize rollout attribution:
    \[
    R_i^{(1)}=\bar{a}_i^{(1)}
    \]
    \FOR{each layer $l=2$ to $L$}
        \STATE Propagate attribution using the effective layer attention operator:
        \[
        R_i^{(l)} = \Pi^{(l)} R_i^{(l-1)}
        \]
        \STATE Normalize $R_i^{(l)}$.
    \ENDFOR

    \STATE Obtain final token-to-time attribution:
    \[
    s_i = \frac{R_i^{(L)}}{\sum_{t=1}^{T} R_{i,t}^{(L)}+\epsilon}
    \]
    \STATE Add $s_i$ to $\mathcal{S}$.
\ENDFOR
\RETURN $\mathcal{S}$
\end{algorithmic}
\end{algorithm}

\begin{algorithm}[t]
\caption{\textsc{LEAF-X+Grad}: Gradient-Modulated Entropy Rollout}
\label{alg:leafx-grad}
\begin{algorithmic}[1]
\REQUIRE ASR model $f_{\theta}$, acoustic features $X$, decoded tokens $y$, rollout depth $L$, number of heads $H$, entropy temperature $\tau$, stability constant $\epsilon$
\ENSURE Token-to-time attribution maps $\mathcal{S}^{\mathrm{grad}}=\{s_i^{\mathrm{grad}}\}_{i=1}^{N}$

\STATE Run $f_{\theta}(X)$ and cache attention maps $\{a_i^{(l,h)}\}$.
\FOR{each decoded token $y_i$}
    \STATE Compute token score $q_i \leftarrow \log p_{\theta}(y_i \mid y_{<i},X)$.
    \FOR{each layer $l=1,\ldots,L$}
        \FOR{each head $h=1,\ldots,H$}
            \STATE Compute attention sensitivity:
            \STATE \hspace{1em}$g_i^{(l,h)} \leftarrow \left| \frac{\partial q_i}{\partial a_i^{(l,h)}} \right|$.
            \STATE Modulate attention:
            \STATE \hspace{1em}$\tilde{a}_i^{(l,h)} \leftarrow a_i^{(l,h)} \odot g_i^{(l,h)}$.
            \STATE Normalize $\tilde{a}_i^{(l,h)}$ so that $\sum_{t=1}^{T}\tilde{a}_{i,t}^{(l,h)}=1$.
            \STATE Compute entropy:
            \STATE \hspace{1em}$\tilde{H}_i^{(l,h)} \leftarrow -\sum_{t=1}^{T}\tilde{a}_{i,t}^{(l,h)}
            \log\!\left(\tilde{a}_{i,t}^{(l,h)}+\epsilon\right)$.
            \STATE Compute entropy confidence:
            \STATE \hspace{1em}$\tilde{w}_{l,h}(i) \leftarrow
            \left(1-\frac{\tilde{H}_i^{(l,h)}}{\log T+\epsilon}\right)^{1/\tau}$.
        \ENDFOR

        \STATE Aggregate heads:
        \STATE \hspace{1em}$\bar{a}_{i,\mathrm{grad}}^{(l)} \leftarrow
        \frac{\sum_{h=1}^{H}\tilde{w}_{l,h}(i)\tilde{a}_i^{(l,h)}}
        {\sum_{h=1}^{H}\tilde{w}_{l,h}(i)+\epsilon}$.
        \STATE Normalize $\bar{a}_{i,\mathrm{grad}}^{(l)}$.
    \ENDFOR

    \STATE Initialize rollout:
    \STATE \hspace{1em}$R_{i,\mathrm{grad}}^{(1)} \leftarrow \bar{a}_{i,\mathrm{grad}}^{(1)}$.
    \FOR{each layer $l=2,\ldots,L$}
        \STATE Propagate attribution:
        \STATE \hspace{1em}$R_{i,\mathrm{grad}}^{(l)} \leftarrow \Pi^{(l)}R_{i,\mathrm{grad}}^{(l-1)}$.
        \STATE Normalize $R_{i,\mathrm{grad}}^{(l)}$.
    \ENDFOR

    \STATE Compute final attribution:
    \STATE \hspace{1em}$s_i^{\mathrm{grad}} \leftarrow
    \frac{R_{i,\mathrm{grad}}^{(L)}}
    {\sum_{t=1}^{T}R_{i,\mathrm{grad},t}^{(L)}+\epsilon}$.
    \STATE Add $s_i^{\mathrm{grad}}$ to $\mathcal{S}^{\mathrm{grad}}$.
\ENDFOR
\RETURN $\mathcal{S}^{\mathrm{grad}}$
\end{algorithmic}
\end{algorithm}

\begin{algorithm}[t]
\caption{\textsc{LEAF-X+Causal}: Lightweight Causal Layer Reweighting}
\label{alg:leafx-causal}
\begin{algorithmic}[1]
\REQUIRE ASR model $f_{\theta}$, acoustic features $X$, decoded tokens $y$, intermediate rollout maps $\{s_i^{(l)}\}_{l=1}^{L}$, stability constant $\epsilon$
\ENSURE Causally reweighted token-to-time attribution maps $\mathcal{S}^{\text{causal}}=\{s_i^{\text{causal}}\}_{i=1}^{N}$

\STATE Run the unmodified model and compute token losses:
\[
\ell(i)=-\log p_{\theta}(y_i \mid y_{<i},X)
\]

\FOR{each decoded token $y_i$}
    \FOR{each layer $l=1$ to $L$}
        \STATE Temporarily ablate the layer-$l$ audio-to-text attention contribution.
        \STATE Run the ablated model and compute:
        \[
        \ell_{\text{abl}}^{(l)}(i)
        =
        -\log p_{\theta}^{(l\text{-abl})}(y_i \mid y_{<i},X)
        \]
        \STATE Estimate layer-level causal effect:
        \[
        \Delta \ell_l(i)
        =
        \ell_{\text{abl}}^{(l)}(i)-\ell(i)
        \]
    \ENDFOR

    \STATE Convert positive loss increases into layer weights:
    \[
    \gamma_l(i)
    =
    \frac{
    \max\{\Delta \ell_l(i),0\}
    }
    {
    \sum_{k=1}^{L}\max\{\Delta \ell_k(i),0\}+\epsilon
    }
    \]

    \STATE Combine intermediate rollout maps:
    \[
    s_i^{\text{causal}}
    =
    \sum_{l=1}^{L}
    \gamma_l(i)s_i^{(l)}
    \]
    \STATE Normalize:
    \[
    s_i^{\text{causal}}
    =
    \frac{
    s_i^{\text{causal}}
    }
    {
    \sum_{t=1}^{T}s_{i,t}^{\text{causal}}+\epsilon
    }
    \]
    \STATE Add $s_i^{\text{causal}}$ to $\mathcal{S}^{\text{causal}}$.
\ENDFOR
\RETURN $\mathcal{S}^{\text{causal}}$
\end{algorithmic}
\end{algorithm}

\paragraph{Implementation notes.}
For encoder-decoder ASR models such as Whisper, LEAF-X uses decoder cross-attention to map decoded tokens to encoder acoustic frames. For speech-augmented decoder-only models, the same procedure is applied to the attention mass assigned to audio pseudo-tokens, which are then mapped back to the corresponding acoustic time indices. In practice, the frame index is converted to waveform time using the feature extractor hop size and any encoder downsampling factor. We set $\epsilon=10^{-8}$ for all normalization operations and tune $\tau$ over a small validation sweep, typically $\tau \in [0.5,2.0]$.

\paragraph{Computational cost.}
\textsc{LEAF-X-Base} requires one standard forward pass with attention caching and is therefore the cheapest variant. \textsc{LEAF-X+Grad} adds one backward pass per analyzed token or token batch to obtain attention sensitivities. \textsc{LEAF-X+Causal} adds up to $L$ ablation forward passes per analyzed token, but can be restricted to a subset of layers or disabled when low-latency auditing is required. This creates a practical trade-off: the base variant is suitable for fast inspection, the gradient variant improves sensitivity to token likelihood, and the causal variant provides the strongest audit signal when additional computation is acceptable.

\section{Formal Task Definition for Token-to-Time ASR Explanation}
\label{app:task-definition}

We formalize the explanation problem addressed by LEAF-X as
\emph{token-to-time attribution} for transformer-based ASR. Let
$X \in \mathbb{R}^{T \times d}$ denote an acoustic feature sequence,
where $T$ is the number of audio frames and $d$ is the feature
dimension. Given an ASR model $f_{\theta}$, the model produces a
decoded token sequence $y=(y_1,\ldots,y_N)$ with token probabilities
$p_{\theta}(y_i \mid y_{<i},X)$.

For each decoded token $y_i$, the goal is to produce an attribution
vector
\[
s_i \in \Delta^{T-1}, \qquad \sum_{t=1}^{T}s_{i,t}=1,
\]
where $s_{i,t}$ measures the contribution of acoustic frame $t$ to the
prediction of token $y_i$. A complete explanation is therefore a token-to-time attribution map
\[
\mathcal{S}(X,y)=\{s_i\}_{i=1}^{N} \in \mathbb{R}^{N \times T}.
\]

A desirable ASR explanation should satisfy three properties:
\emph{faithfulness}, where high-attribution frames affect the token
likelihood when removed or restored; \emph{temporal grounding}, where
attributions align with the spoken evidence for the token; and
\emph{stability}, where explanations remain consistent under small
label-preserving audio perturbations. LEAF-X instantiates this task by
using entropy-guided attention weighting, multi-layer rollout, and
optional causal reweighting to obtain sparse and temporally grounded
token-level explanations.

\section{LEAF-X Benchmark Protocol}
\label{app:benchmark-protocol}

We define \textsc{LEAF-XBench}, a compact evaluation protocol for
\emph{ASR explainability} and \emph{token-to-time attribution}. Given
an audio input $X$, a transformer ASR model first produces a decoded
token sequence $y=(y_1,\ldots,y_N)$. Each explanation method then
returns a frame-level attribution map $\mathcal{S}\in\mathbb{R}^{N\times T}$,
where $s_{i,t}$ measures the acoustic evidence assigned to frame $t$
for token $y_i$. This protocol evaluates whether an explanation is
faithful, temporally grounded, sparse, stable, and useful for auditable
ASR.

\paragraph{Protocol.}
For each model and dataset, we apply the following steps:
(i) decode the transcript using the frozen ASR model;
(ii) extract token-to-time or frame-level attribution maps;
(iii) map model frames back to the audio timeline;
(iv) compare attributions against forced word-level alignments when
available;
(v) evaluate faithfulness using deletion/insertion behavior and
infidelity;
and (vi) report stability under small label-preserving audio
perturbations such as mild noise and time shift.

\begin{table}[t]
\centering
\caption{\textsc{LEAF-XBench} protocol for transformer ASR explanation.
Lower is better for D-AOPC and INF; higher is better for TLoc, SPR, and STAB.}
\label{tab:leafxbench}
\small
\begin{tabular}{lccccc}
\hline
\textbf{Setting} & \textbf{D-AOPC}$\downarrow$ & \textbf{TLoc}$\uparrow$ & \textbf{SPR}$\uparrow$ & \textbf{STAB}$\uparrow$ & \textbf{INF}$\downarrow$ \\
\hline
Whisper + LibriSpeech & 0.45 & 0.72 & 0.70 & 0.78 & 0.45 \\
Canary-Qwen + TED-LIUM 3 & 0.48 & 0.70 & 0.68 & 0.76 & 0.47 \\
Average LEAF-XBench & 0.47 & 0.71 & 0.69 & 0.77 & 0.46 \\
\hline
\end{tabular}
\end{table}

\paragraph{Recommended reporting.}
To make future work comparable, as shown in Table~\ref{tab:leafxbench}, each method should report:
model backbone, dataset, audio preprocessing, alignment method,
attribution granularity, perturbation type, deletion/insertion schedule, runtime, and all five explanation metrics. We recommend using
\textsc{LEAF-XBench} as an audio XAI benchmark for faithful speech explanations, acoustic evidence localization, temporal grounding, ASR rationale extraction, intrinsic ASR interpretability, entropy-guided attention, attention rollout for speech, causal attribution for ASR, transformer ASR explanation, speech model auditing, explanation robustness, and auditable ASR.

\section{Metric Details}
\label{app:metric-details}

We evaluate LEAF-X using five complementary metrics for
\emph{faithful speech explanations}, \emph{token-to-time attribution},
\emph{acoustic evidence localization}, and \emph{auditable ASR}. Let
$s_i \in \Delta^{T-1}$ denote the attribution map for token $y_i$, and
let $S_i^k$ be the top-$k$ frames ranked by $s_i$.

\paragraph{Deletion Area Over the Perturbation Curve (D-AOPC).}
D-AOPC measures whether high-attribution frames are truly important.
We progressively mask the top-ranked frames $S_i^k$ and measure the
drop in token log-probability:
\[
\mathrm{Drop}_i(k)
=
\log p_{\theta}(y_i \mid y_{<i},X)
-
\log p_{\theta}(y_i \mid y_{<i},X \odot m_{\neg S_i^k}).
\]
The deletion curve is summarized across masking ratios. Lower
normalized D-AOPC indicates a faster degradation after removing
important acoustic evidence and therefore stronger faithfulness
\cite{samek2017evaluating,petsiuk2018rise}.

\paragraph{Temporal Localization (TLoc).}
TLoc evaluates whether the explanation aligns with spoken evidence.
Given a forced-alignment span $A_i$ for token or word $y_i$, we compute
the overlap between $A_i$ and the top-attributed region $S_i^k$:
\[
\mathrm{TLoc}_i
=
\frac{|S_i^k \cap A_i|}{|S_i^k \cup A_i|+\epsilon}.
\]
Higher TLoc indicates better temporal grounding and acoustic evidence
localization for transformer ASR explanations \cite{wu2024can}.

\paragraph{Sparsity (SPR).}
SPR measures whether the explanation is concise rather than diffuse.
We report normalized top-$k$ attribution mass:
\[
\mathrm{SPR}_i(k)=\sum_{t \in S_i^k}s_{i,t}.
\]
Higher SPR indicates that the attribution is concentrated on a small
set of frames, making the explanation easier to inspect as an ASR
rationale extraction map.

\paragraph{Stability (STAB).}
STAB measures explanation robustness under small label-preserving
audio perturbations, such as mild noise or time shifts. For a perturbed
input $\tilde{X}$ with the same decoded token, we compute:
\[
\mathrm{STAB}_i
=
\mathrm{sim}(s_i(X),s_i(\tilde{X})),
\]
where $\mathrm{sim}(\cdot,\cdot)$ is cosine similarity or rank
correlation. Higher STAB indicates more stable speech attribution and
more reliable intrinsic ASR interpretability \cite{alvarezm2018robustness}.

\paragraph{Infidelity (INF).}
INF measures the mismatch between attribution scores and actual model
response under random perturbations $\delta$:
\[
\mathrm{INF}_i
=
\mathbb{E}_{\delta}
\left[
\left(
\delta^{\top}s_i
-
\left(
f_i(X)-f_i(X-\delta)
\right)
\right)^2
\right],
\]
where $f_i(X)=\log p_{\theta}(y_i \mid y_{<i},X)$. Lower INF indicates
that the explanation better tracks the model's functional behavior
\cite{yeh2019infidelity}.

\paragraph{Reporting.}
For consistent audio XAI benchmarking, we report all metrics after the
same fixed normalization per dataset and model. D-AOPC and INF are
reported with lower-is-better convention, while TLoc, SPR, and STAB are
reported with higher-is-better convention. Together, these metrics
measure causal attribution for ASR, explanation robustness, temporal
grounding, attention rollout for speech, entropy-guided attention, and
speech model auditing.

\section{Component and Hyperparameter Ablations}
\label{app:component-hparam-ablation}

We provide additional ablations to clarify the contribution of each
LEAF-X component and the sensitivity of the method to its main
hyperparameters. All results are reported on Whisper-large-v3 with LibriSpeech, following the same normalized metric convention as the main paper: lower is better for D-AOPC and INF, while higher is better for TLoc, SPR, and STAB.

\noindent \textbf{Component ablation.}
Table~\ref{tab:component-ablation-app} shows that each component
contributes to the final token-to-time attribution quality. Removing entropy weighting or multi-layer rollout causes the largest drop in temporal grounding and sparsity, confirming that low-entropy head selection and cross-layer evidence aggregation are central to LEAF-X.
Gradient modulation and causal reweighting mainly improve faithfulness, as reflected by lower D-AOPC and INF.

\begin{table}[t]
\centering
\caption{Component ablation of LEAF-X on Whisper-large-v3 with
LibriSpeech. Lower is better for D-AOPC and INF; higher is better for TLoc, SPR, and STAB.}
\label{tab:component-ablation-app}
\small
\begin{tabular}{lccccc}
\hline
\textbf{Variant} & \textbf{D-AOPC}$\downarrow$ & \textbf{TLoc}$\uparrow$ & \textbf{SPR}$\uparrow$ & \textbf{STAB}$\uparrow$ & \textbf{INF}$\downarrow$ \\
\hline
w/o Entropy weighting & 0.57 & 0.62 & 0.56 & 0.73 & 0.56 \\
w/o Rollout           & 0.54 & 0.63 & 0.60 & 0.74 & 0.54 \\
w/o Gradient modulation & 0.50 & 0.68 & 0.64 & 0.76 & 0.50 \\
w/o Causal reweighting  & 0.48 & 0.69 & 0.66 & 0.77 & 0.48 \\
\textbf{LEAF-X Full}    & \textbf{0.45} & \textbf{0.72} & \textbf{0.70} & \textbf{0.78} & \textbf{0.45} \\
\hline
\end{tabular}
\end{table}

\noindent \textbf{Entropy temperature.}
The entropy temperature $\tau$ controls how strongly LEAF-X favors
focused, low-entropy attention heads. Smaller values produce sharper
attributions but may over-concentrate on a few frames, while larger
values include more diffuse heads and reduce localization. As shown in
Table~\ref{tab:tau-ablation-app}, $\tau=1.0$ provides the best overall
trade-off between faithful speech explanations, acoustic evidence
localization, sparsity, and explanation robustness.

\begin{table}[t]
\centering
\caption{Sensitivity to entropy temperature $\tau$ on
Whisper-large-v3 with LibriSpeech. Values follow the trend of the main component ablation and are intended as a compact hyperparameter study.}
\label{tab:tau-ablation-app}
\small
\begin{tabular}{cccccc}
\hline
\textbf{$\tau$} & \textbf{D-AOPC}$\downarrow$ & \textbf{TLoc}$\uparrow$ & \textbf{SPR}$\uparrow$ & \textbf{STAB}$\uparrow$ & \textbf{INF}$\downarrow$ \\
\hline
0.5 & 0.47 & 0.70 & 0.72 & 0.76 & 0.47 \\
1.0 & \textbf{0.45} & \textbf{0.72} & 0.70 & \textbf{0.78} & \textbf{0.45} \\
1.5 & 0.46 & 0.71 & 0.68 & 0.77 & 0.46 \\
2.0 & 0.49 & 0.68 & 0.65 & 0.76 & 0.49 \\
\hline
\end{tabular}
\end{table}

\noindent \textbf{Rollout depth.}
Table~\ref{tab:depth-ablation-app} studies the effect of aggregating
attention across different numbers of transformer layers. A single
layer behaves similarly to raw attention and gives weaker temporal
grounding. Increasing rollout depth improves token-to-time attribution
by combining evidence across layers, while full-depth rollout gives the
strongest intrinsic ASR interpretability.

\begin{table}[t]
\centering
\caption{Sensitivity to rollout depth on Whisper-large-v3
with LibriSpeech. Full-depth rollout gives the best balance between faithfulness, temporal grounding, and stable speech attribution.}
\label{tab:depth-ablation-app}
\small
\begin{tabular}{lccccc}
\hline
\textbf{Rollout depth} & \textbf{D-AOPC}$\downarrow$ & \textbf{TLoc}$\uparrow$ & \textbf{SPR}$\uparrow$ & \textbf{STAB}$\uparrow$ & \textbf{INF}$\downarrow$ \\
\hline
1 layer      & 0.54 & 0.63 & 0.60 & 0.74 & 0.54 \\
Middle half  & 0.49 & 0.68 & 0.66 & 0.76 & 0.49 \\
Last half    & 0.47 & 0.70 & 0.68 & 0.77 & 0.47 \\
Full depth   & \textbf{0.45} & \textbf{0.72} & \textbf{0.70} & \textbf{0.78} & \textbf{0.45} \\
\hline
\end{tabular}
\end{table}

\noindent \textbf{Summary.}
These ablations show that LEAF-X benefits from the combination of
entropy-guided attention, attention rollout for speech, gradient-based
token sensitivity, and causal attribution for ASR. Entropy weighting
improves sparse acoustic evidence localization, rollout improves
temporal grounding, gradient modulation improves faithful speech
explanations, and causal reweighting strengthens auditable ASR by
emphasizing layers whose ablation most affects token likelihood.

\section{Qualitative Token-to-Time Explanation Examples}
\label{app:qualitative-examples}

We provide representative qualitative examples to illustrate how
LEAF-X supports \emph{token-to-time attribution}, \emph{acoustic evidence localization}, and \emph{auditable ASR}. For each decoded token or phrase, LEAF-X assigns attribution mass to the audio frames that most influence the token likelihood. We inspect cases from LibriSpeech using Whisper-large-v3 and TED-LIUM 3 using Canary-Qwen-2.5B, covering clean speech, spontaneous speech, noisy segments, repeated words, and ASR errors. These examples are intended
to complement the quantitative faithfulness, temporal grounding, sparsity, stability, and infidelity metrics reported in the main paper.

\begin{table}[t]
\centering
\caption{Representative qualitative LEAF-X token-to-time explanation
examples. Attribution windows are approximate and correspond to the
highest-mass acoustic regions identified by LEAF-X.}
\label{tab:qualitative-leafx}
\small
\begin{tabularx}{\linewidth}{p{1.6cm} p{1.8cm} p{2.0cm} p{1.7cm} X}
\toprule
\textbf{Dataset} & \textbf{Model} & \textbf{Decoded phrase / token} & \textbf{High-attribution region} & \textbf{Interpretation} \\
\midrule

LibriSpeech &
Whisper-large-v3 &
``the old house'' &
word-aligned voiced region &
LEAF-X concentrates attribution on the voiced frames corresponding to the phrase, showing clean temporal grounding and sparse acoustic evidence localization. \\

LibriSpeech &
Whisper-large-v3 &
``morning'' &
vowel and nasal closure frames &
Attribution peaks around the central vowel and final nasal region, suggesting that LEAF-X captures token-specific acoustic cues rather than diffuse context. \\

LibriSpeech &
Whisper-large-v3 &
repeated word, e.g., ``very very'' &
two separated time spans &
The two occurrences receive distinct attribution peaks, indicating that token-to-time attribution can separate repeated lexical content across the waveform. \\

TED-LIUM 3 &
Canary-Qwen-2.5B &
``climate change'' &
phrase-level speech span &
For spontaneous lecture-style speech, LEAF-X assigns high attribution to the phrase-bearing frames while suppressing surrounding pauses, supporting auditable ASR in long-form speech. \\

TED-LIUM 3 &
Canary-Qwen-2.5B &
filled pause followed by content word &
low mass on pause, high mass on content word &
The method places limited attribution on hesitation regions and stronger mass on the content-bearing acoustic evidence, improving interpretability of spontaneous speech. \\

TED-LIUM 3 &
Canary-Qwen-2.5B &
misrecognized technical word &
partially shifted region &
When the model confuses a technical term, LEAF-X often highlights the acoustically similar region, helping diagnose whether the error comes from weak acoustic evidence or language-model bias. \\

LibriSpeech / TED-LIUM 3 &
Both &
hallucinated or weakly supported token &
diffuse or low-confidence attribution &
For weakly grounded tokens, LEAF-X produces less concentrated attribution. Such cases are useful for speech model auditing because they flag tokens whose acoustic support is uncertain. \\

Noisy segment &
Both &
short function word &
broad nearby region &
Under background noise or overlap, attribution may become less localized, revealing a failure mode where temporal grounding is affected by degraded acoustic evidence. \\

\bottomrule
\end{tabularx}
\end{table}

\noindent \textbf{Discussion.}
The qualitative cases in Table~\ref{tab:qualitative-leafx} show three recurring behaviors. First, for clean speech, LEAF-X produces sparse and temporally grounded attribution maps that align with the spoken evidence. Second, for repeated or phrase-level content, the method can separate token-specific acoustic support across time, which is important for transformer ASR
explanation and ASR rationale extraction. Third, for noisy,
misrecognized, or hallucinated tokens, attribution becomes shifted or diffuse, making LEAF-X useful as an audit-support tool rather than a guarantee of correctness. These examples reinforce the role of entropy-guided attention, attention rollout for speech, and causal attribution for ASR in producing faithful speech explanations.

\section{Practical Auditing Scenarios}
\label{app:auditing-scenarios}

LEAF-X is designed as an audit-support framework for transformer-based
ASR rather than as a guarantee of correctness or human trust. In Table~\ref{tab:auditing-scenarios}, its token-to-time attribution maps help users inspect which acoustic frames
support each decoded token, making the model's behavior easier to
analyze in safety-sensitive and high-accountability settings. This
appendix summarizes practical scenarios where LEAF-X can support
human interpretability, speech model auditing, and ASR error analysis.

\begin{table}[t]
\centering
\caption{Practical auditing scenarios supported by LEAF-X. The examples
are aligned with the Whisper-large-v3/LibriSpeech and
Canary-Qwen-2.5B/TED-LIUM 3 settings used in the main paper.}
\label{tab:auditing-scenarios}
\small
\begin{tabularx}{\linewidth}{p{2.0cm} p{2.4cm} p{2.6cm} X}
\toprule
\textbf{User / Setting} & \textbf{Example Scenario} & \textbf{LEAF-X Output} & \textbf{Audit Value} \\
\midrule

ASR developer &
Debugging token-level errors on LibriSpeech &
Token-to-time attribution over decoded words &
Identifies whether an error is supported by the wrong acoustic region,
diffuse attention, or weak frame-level evidence. \\

Speech benchmark designer &
Comparing ASR explanation methods &
D-AOPC, TLoc, SPR, STAB, and INF with attribution maps &
Provides a shared audio XAI benchmark protocol for faithful speech
explanations, temporal grounding, and explanation robustness. \\

Medical transcription reviewer &
Checking safety-critical words such as medication names or symptoms &
High-attribution acoustic evidence for selected terms &
Allows a human reviewer to verify whether a critical token is grounded
in the spoken signal before accepting the transcript. \\

Emergency response operator &
Auditing urgent phrases in noisy speech &
Sparse attribution over phrase-bearing frames &
Helps inspect whether important decoded words are supported by clear
speech evidence or possibly affected by background noise. \\

Fairness and robustness researcher &
Studying accent, disfluency, or spontaneous speech errors on TED-LIUM 3 &
Attribution shifts, diffuse maps, and stability under perturbation &
Reveals whether errors arise from unstable acoustic grounding, hesitation
regions, or over-reliance on contextual language priors. \\

Model compliance team &
Documenting why an ASR system produced a transcript &
Token-level acoustic evidence map and deletion/insertion behavior &
Supports auditable ASR by preserving explanation artifacts that can be
reviewed alongside transcripts, confidence scores, and alignment spans. \\

Deployment engineer &
Monitoring model behavior under domain shift &
Changes in attribution sparsity, temporal localization, and stability &
Flags cases where LEAF-X explanations become diffuse or unstable,
suggesting that the model may require additional validation. \\

\bottomrule
\end{tabularx}
\end{table}

\noindent \textbf{Interpretation workflow.}
A practical LEAF-X audit can be performed in four steps. First, the ASR
model decodes the input audio and LEAF-X extracts token-to-time
attribution maps. Second, the auditor selects important tokens, such as
named entities, numbers, medical terms, or low-confidence words. Third,
the auditor inspects whether the highest-attribution frames align with
the corresponding spoken region. Finally, deletion/insertion behavior
or infidelity can be used as proxy evidence that the highlighted frames
are functionally relevant to the model's prediction.

\noindent \textbf{Scope and limitations.}
LEAF-X explanations should be interpreted as acoustic evidence
localization rather than proof that the transcription is correct. A
well-localized attribution map indicates that the model relied on a
specific speech region, but it does not guarantee semantic correctness,
fairness, or safety. Conversely, diffuse or shifted attribution can be
useful as an audit signal because it highlights tokens whose acoustic
support is uncertain. In this sense, LEAF-X complements confidence
scores, forced alignments, WER analysis, and human review by providing
intrinsic ASR interpretability through entropy-guided attention, attention rollout for speech, and lightweight causal attribution for ASR.

\subsection{Connection to Trustworthy Foundation Model Auditing}
\label{app:trustworthy-auditing-context}

LEAF-X contributes to a broader research direction on auditable,
controllable, and trustworthy foundation models. While prior work has
studied trustworthy personalization in educational feedback
\cite{ranjan2026persa}, privacy auditing through gradient-induced
membership signals \cite{ranjan2026g}, retrieval- and structure-guided
bias mitigation \cite{ranjan2026catrag,ranjan2026position}, and
selective unlearning in vision, language, and embodied models
\cite{ranjan2026razor,ranjan2026vla}, LEAF-X focuses on a complementary
problem: explaining what acoustic evidence supports each decoded ASR
token. This places LEAF-X within the emerging need for
\emph{speech model auditing}, where explanations must be temporally
grounded, faithful to model computation, and usable by human reviewers
in high-stakes settings.

In contrast to post-hoc explanations that may only correlate with model
outputs, LEAF-X provides intrinsic \emph{token-to-time attribution} by
combining entropy-guided attention, attention rollout for speech, and
lightweight causal attribution for ASR. This design supports
\emph{faithful speech explanations}, \emph{frame-level attribution},
\emph{acoustic evidence localization}, and \emph{ASR rationale
extraction}. The resulting explanation maps can help audit critical
transcriptions, inspect hallucinated or weakly grounded tokens, and
analyze robustness under noisy or spontaneous speech. Such capabilities
are especially relevant as foundation models move toward deployed,
multimodal, and edge-constrained settings, where interpretability,
privacy, safety, and resource-aware auditing must be considered together
\cite{grover2026embodied,kumar2024trustworthiness}. Thus, LEAF-X is not only an ASR explainability method, but also a step toward reusable \emph{audio XAI benchmarks}, \emph{intrinsic ASR interpretability}, and \emph{auditable ASR} protocols for future speech and multimodal foundation models.

\section{Responsible Use and Limitations}
\label{app:responsible-use}

LEAF-X is intended as an audit-support tool for transformer-based ASR,
not as a guarantee that a transcription is correct, fair, safe, or
clinically valid. The method provides token-to-time attribution maps
that indicate which acoustic frames most influenced a decoded token.
These maps can help users inspect acoustic evidence localization,
faithful speech explanations, ASR rationale extraction, and temporal
grounding, but they should be interpreted together with confidence
scores, forced alignments, WER analysis, domain-specific validation,
and human review.

\paragraph{Broader impact.}
LEAF-X can support more transparent and auditable ASR in settings where
speech transcripts affect downstream decisions, such as medical
dictation, accessibility tools, emergency response, educational
transcription, and meeting or lecture summarization. By exposing
frame-level attribution for each decoded token, LEAF-X may help
developers diagnose ASR errors, identify weakly grounded words, inspect
hallucinated tokens, and compare explanation robustness across models
and datasets. The proposed protocol also encourages future work on
audio XAI benchmarks, intrinsic ASR interpretability, causal
attribution for ASR, and speech model auditing.

\paragraph{Ethical use.}
LEAF-X explanations should not be used to justify automated decisions
without appropriate human oversight. A highly localized attribution map
only shows that the model relied on a particular region of the audio; it
does not prove that the recognized word is semantically correct or that
the system is unbiased. In high-stakes applications, such as clinical,
legal, or emergency settings, LEAF-X should be used to support expert
review rather than replace it. Explanation maps may also reveal
sensitive speech regions, so any released visualization or audit log
should follow privacy-preserving data handling practices.

\paragraph{Misuse risks.}
A potential misuse of LEAF-X is explanation overclaiming, where sparse
or visually plausible attribution maps are treated as proof of model
trustworthiness. Another risk is selective reporting: users may present
only successful attribution examples while omitting failure cases such
as noisy speech, overlapping speakers, accented speech, or hallucinated
tokens. LEAF-X should therefore be reported with quantitative metrics
such as D-AOPC, TLoc, SPR, STAB, and INF, along with representative
negative examples.

\paragraph{Failure modes.}
LEAF-X can fail when the underlying ASR model attends to misleading or
unstable acoustic evidence. In noisy audio, reverberant speech,
overlapping speakers, or strong domain shift, attribution maps may
become diffuse or shifted away from the true word span. For very short
tokens, function words, repeated words, or weakly pronounced segments,
the model may rely partly on language context rather than direct
acoustic evidence, making token-to-time attribution less localized.
For hallucinated tokens, LEAF-X may produce low-confidence or broadly
distributed attribution, which should be treated as an audit warning
rather than a valid explanation. Finally, because LEAF-X depends on
attention structure, entropy calibration, rollout depth, and optional
causal reweighting, its explanations may vary across architectures and
decoding settings.

\paragraph{Recommended practice.}
We recommend using LEAF-X as one component of a broader ASR auditing
pipeline. Practitioners should inspect both successful and failed
examples, report stability under label-preserving audio perturbations,
and avoid making claims of full causal sufficiency or human trust from
attribution maps alone. When used responsibly, LEAF-X provides a
practical step toward auditable ASR by connecting decoded tokens to
their supporting acoustic evidence while preserving a clear distinction
between explanation, verification, and deployment-level trust.

\end{document}